\documentclass[a4paper,11pt,hyphens]{amsart} 

\usepackage{bbm}
\usepackage{enumitem}
\usepackage{amsrefs}
\usepackage{multirow}
\usepackage{graphicx}
\usepackage{tikz}
\usetikzlibrary{positioning}
\usepackage{MnSymbol}

\usepackage[hidelinks]{hyperref}
\hypersetup{breaklinks=true}
\usepackage{cite}
\usepackage[hyphenbreaks]{breakurl}

\tikzset{basic/.style={draw,fill=white!20,text width=1em,text badly centered}}
\tikzset{input/.style={basic,rectangle, text width=12em}}
\tikzset{weights/.style={basic,rectangle}}
\tikzset{functions/.style={basic,circle,fill=white!10}}

\title{Quantum Cryptography: Quantum Key Distribution, a Non-technical Approach}
\author{Andrew B. Frigyik}
\date{\today}

\address{\'Obuda University, Don\'at B\'anki Faculty of Mechanical and Safety
  Engineering, Institute of Mechatronics \& Vehicle Engineering, Dept. of
  Mechatronics, Budapest, Hungary} \email{frigyik.andras@bgk.uni-obuda.hu}

\begin{document}

\begin{abstract}
  With the rapid development of quantum computers the currently secure
  cryptographic protocols may not stay that way. Quantum mechanics provides
  means to create an inherently secure communication channel that is protected
  by the laws of physics and not by the computational hardness of certain
  mathematical problems. This paper is a non-technical overview of quantum key
  distribution, one of the most well-known application of quantum cryptography,
  a type of cryptography poised to exploit the laws of quantum mechanics
  directly.
\end{abstract}

\maketitle

\section{
Introduction
}

Q-Day is coming! In a recently published {\it Feature} article in {\it Nature}
\cite{castelvecchi2022race} the author assesses the threat Quantum computers
impose on the cryptographic systems currently in use. The aforementioned Q-Day
refers to the day when the quantum computers will have become powerful enough
to break the security schemes of the day.

The safety of our communication systems today is based on mathematical problems
that are hard: Even if we had an access to a super computer we
won't be able to do the required calculations in reasonable time. Prime
factorisation (splitting a number into prime factors) or the problem of discrete
logarithm (a number theoretic version of the well-known problem of finding the
logarithm of a number) are hard because there are no quick or efficient
algorithms to solve them. On the other hand, if someone claims to know the
result for such a computation, it is easy to check whether that person is
correct.

If the proponents of quantum computing are right the situation can change
overnight. Quantum computers promise to be able to run algorithms that can solve
current essentially unsolvable problems very efficiently. In order to keep our
data secure there are several strategies we can implement. One way to go is
post-quantum cryptography. Another approach is to make the delivery of the key that
is required for the decryption extremely secure: We could enroll quantum
mechanics to make the communication channel secure against any kind of
eavesdropping. If the eavesdropper has nothing to decode then it doesn't matter
what kind of computer they have access to. The aim of this article is to give a non-technical survey of one
of the contemporary methods: the quantum key distribution. I hope to shed some
light on the role quantum mechanics is playing in the process.

The paper is organized as follows: Section 2 is about the very first system that
utilised quantum mechanics to secure a communication channel. This system is
based solely on the Heisenberg principle. Section 3 describes systems that use
entanglement, another feature of quantum mechanics, to make channels secure. It
is not easy to create and control the quantum systems necessary for successful
operation. In general, complex systems are more vulnerable than simple ones. The
device independent realisations of the protocols get rid of this
disadvantage: Their successful deployment does not depend on the inner workings
of a device, only on the fact that it functions as prescribed. Some solutions can even handle the case when the hardware was produced
by an adversarial party. Section 4 covers some of the latest developments in
this area.

In this paper I would like to concentrate on the quantum aspect of
cryptography. For classical cryptographic terminology please refer to any
of the many textbooks available (e.g.,
\cite{paar2009understanding}). There are a lot of technical surveys on quantum
key distribution, e.g. \cite{bruss2007quantum} and \cite{kumar2021state}, just to mention a few.

The idea of this paper came to me while I was reading Mordechai Rorvig's
article \cite{rorvig2022crypto} in {\it Quanta magazine}  and the paper \cite{nyari2021impact} by Norbert Ny\'ari
published in {\it Safety and Security Sciences Review (Biztons\'agtudomanyi Szemle)}. 

\section{Basic Idea: BB84}

The goal of key distribution in cryptography is to share, in a very secure
manner (no expense is spared), a small amount of information in order to use it
as a one-time pad to code and decode much more information later, still
securely. It is assumed that the communicating parties did not share any secret
information beforehand: They have to rely on the presently shared key if they
want communicate securely.

In case
of quantum key distribution, the communicating parties recruit quantum mechanics to
make sure their key is distributed safely. The first paper on quantum key distribution was published by Charles H. Bennett
and Gilles Brassard in 1984 \cites{bennett1984quantum,BENNETT20147}, hence the
designation BB84. As usual, we are going to call the communicating parties Alice
and Bob and the eavesdropping adversary will be called Eve. Alice would like to
share a key with Bob, only, in order to communicate with him securely. To do
this she is going to use a quantum and a classical channel, but only when the protocol indicates it. There is no communication between them outside of the protocol. The quantum channel
uses quantum bits, bits that behave observably according to quantum mechanics. The
classical channel uses conventional classical bits. Those two kinds of bits behave very differently
when we ask questions from them.

One way to picture a bit is to consider a dial
with a pointer. There are a lot of different ways to represent the values $1$
and $0$ on such a dial. For example, the pointer could point north (up) or
south (down) and these positions would correspond to the value $1$. Or it could
point to east or to west, corresponding to the value $0$. Another way to
express these values on the dial is to assign the value $1$ to pointers pointing
north-east or south-west and the value $0$ to pointers aiming north-west or
south-east. 

Suppose Alice prepares a classical bit using this representation (the dial and
the pointer) and sends it to Bob. Bob
has two masks: One, called Rectilinear (R), with slits oriented north/south and
east/west. The other one, called Diagonal (D), with slits oriented north-east/south-west and north-west/south-east. If Alice
sent a dial with pointer aimed north and Bob uses the Rectilinear mask, he can
read out the value of the bit, which is $1$. If he uses the Diagonal mask, he
won't get a value. All he learns is the fact that he used the wrong mask. At
least he learns something even in this case. 

Now, suppose Alice prepares a quantum bit using the same kind of representation,
only this time the dial and the pointer are quantum. She, again, sends the bit to Bob. The set of masks Bob possesses is the same as
before. If Alice sent a dial with pointer aimed north and Bob applies the
Rectilinear mask he will get the correct value of $1$ every time. That is, if
Alice keeps sending the same dial prepared in the same way again and again and
Bob keeps applying the same mask again and again, he is going to get the correct
value every time. But if Bob chooses to use the Diagonal mask he will get a
value: It will be $1$ approximately half of the time and $0$ the other half of
the time. This is the consequence of the quantum nature of the bit. If Alice
sends only one quantum bit to Bob and he can choose a mask freely then there is no way for him to know, at this
point, whether he got the right result or used the wrong mask and got a
meaningless result. 

How can this phenomenon help us to achieve our goal of creating an inherently
secure way of communication? One answer is the BB84 protocol. 

The process begins with Alice creating a sequence of $n$ random bits. Part of this
sequence will form the key at the end of the process. She creates another
sequence of $n$ bits and uses this second sequence to choose a
representation for each of the bits in her first sequence. For example, each bit
of value $1$ in her second sequence means she is going to use the Rectilinear
representation (i.e. a representation readable correctly using the Rectilinear
mask) for the corresponding bit in her first sequence. Similarly a
bit of value $0$ in her second sequence will correspond to a Diagonal
representation readable correctly by the Diagonal mask. 
 
Suppose Alice obtained twelve bits on her first run and another twelve on her
second run. Table \ref{first} shows the first steps of the protocol. Alice
decided to use north/south pointer to represent the value $1$ in case of
the Rectilinear representation and the north-east/south-west pointer to
represent the same value in case of the Diagonal representation. 

\begin{table}[ht]
  \caption{First steps in BB84 protocol.}
   \resizebox{\textwidth}{!}
  {%
    \begin{tabular}{lcccccccccccc}
      \hline
Alice's random bits  & 1 & 0 & 1 & 0 & 0 & 1 & 0 & 0 & 0 & 1 & 1 & 1 \\  \hline
Random bits for rep. & 1 & 1 & 0 & 1 & 1 & 0 & 0 & 0 & 1 & 1 & 0 & 0 \\  \hline
Type of rep.         & R & R & D & R & R & D & D & D & R & R & D & D \\  \hline
Actual rep. &
  $\updownarrow$ &
  $\leftrightarrow$ &
  $\neswarrow$ &
  $\leftrightarrow$ &
  $\leftrightarrow$ &
  $\neswarrow$ &
  $\nwsearrow$ &
  $\nwsearrow$ &
  $\leftrightarrow$ &
  $\updownarrow$ &
  $\neswarrow$ &
  $\neswarrow$ \\ \hline
\end{tabular}%
}
\label{first}
\end{table}

Next Alice sends her representations via quantum channel, for example via
fiber-optic cable using photons, to Bob. Since Bob doesn't know what
representation Alice used for the individual bits, he creates twelve (in general
$n$) random bits and chooses masks accordingly. For example, if the bit he got
is $1$ he is going to use the Rectilinear mask and the Diagonal mask
otherwise. Table \ref{second} shows the result of his decision.

\begin{table}[ht]
  \caption{Bob's measurements.}
   \resizebox{\textwidth}{!}
  {%
    \begin{tabular}{lcccccccccccc}
       \hline
Alice's random bits  & 1 & 0 & 1 & 0 & 0 & 1 & 0 & 0 & 0 & 1 & 1 & 1 \\ \hline
Random bits for rep. & 1 & 1 & 0 & 1 & 1 & 0 & 0 & 0 & 1 & 1 & 0 & 0 \\ \hline
Type of rep.         & R & R & D & R & R & D & D & D & R & R & D & D \\ \hline
Actual rep. &
  $\updownarrow$ &
  $\leftrightarrow$ &
  $\neswarrow$ &
  $\leftrightarrow$ &
  $\leftrightarrow$ &
  $\neswarrow$ &
  $\nwsearrow$ &
  $\nwsearrow$ &
  $\leftrightarrow$ &
  $\updownarrow$ &
  $\neswarrow$ &
  $\neswarrow$ \\ \hline
Bob's random bits    & 1 & 1 & 0 & 1 & 1 & 1 & 0 & 0 & 1 & 0 & 0 & 0 \\ \hline
Type of mask         & R & R & D & R & R & R & D & D & R & D & D & D \\ \hline
\end{tabular}%
}
\label{second}
\end{table}

Where the actual representation and the mask type coincide Bob gets the correct
value every time. If there is mismatch the result will be random: about half the
time Bob sees a $1$ and the other half the time a $0$. Once Bob is done with the
inspection the communicating parties use a classical public channel that is
susceptible to eavesdropping but not to any kind of modification of the
communication. They compare the types of representation Alice used with the
masks Bob used. They keep those bits where the mask matches the type (Table \ref{third}).

\begin{table}[ht]
  \caption{Bits to keep.}
\resizebox{\textwidth}{!}{%
\begin{tabular}{lcccccccccccc}
\hline
Alice's random bits  & 1  & 0  & 1  & 0  & 0  & 1 & 0  & 0  & 0  & 1 & 1  & 1  \\ \hline
Random bits for rep. & 1  & 1  & 0  & 1  & 1  & 0 & 0  & 0  & 1  & 1 & 0  & 0  \\ \hline
Type of rep.         & R  & R  & D  & R  & R  & D & D  & D  & R  & R & D  & D  \\ \hline
Actual rep. &
  $\updownarrow$ &
  $\leftrightarrow$ &
  $\neswarrow$ &
  $\leftrightarrow$ &
  $\leftrightarrow$ &
  $\neswarrow$ &
  $\nwsearrow$ &
  $\nwsearrow$ &
  $\leftrightarrow$ &
  $\updownarrow$ &
  $\neswarrow$ &
  $\neswarrow$ \\ \hline
Bob's random bits    & 1  & 1  & 0  & 1  & 1  & 1 & 0  & 0  & 1  & 0 & 0  & 0  \\ \hline
Type of mask         & R  & R  & D  & R  & R  & R & D  & D  & R  & D & D  & D  \\ \hline
Received bits        & 1  & 0  & 1  & 0  & 0  & 0 & 0  & 0  & 0  & 0 & 1  & 1  \\ \hline
Bits to keep         & ok & ok & ok & ok & ok &   & ok & ok & ok &   & ok & ok \\ \hline
\end{tabular}%
}

\label{third}
\end{table}

Suppose Eve, the eavesdropper, is listening in on both the quantum and the
classical channel. Suppose the system designers followed at least one of
Kerckhoffs' principles and made the technical details of the protocol and the
physical implementation public. In this case Eve knows that she needs only two
type of masks and knows what kind of masks they are. Nevertheless, if she
interacts with the quantum bits sent by Alice, she runs the risk of modifying
them: Using the wrong kind of mask does not simply erase the sent bit but
changes the type of representation of it. For Eve, the result is
indistinguishable from a valid result. If Eve could copy the bits that were sent
by Alice then she could listen in on the public channel and learn about the
correct masks and which bits to keep. The No-cloning theorem of quantum
mechanics prevents her to do exactly this. Adapted to this situation, the
theorem says that it is not possible to create an exact copy of an arbitrary
unknown quantum bit. Since for Eve the intercepted quantum bit is unknown,
quantum mechanics prevents her from making an exact copy of it.

Finally, Alice and Bob have to make sure the quantum channel they used was not
compromised. In order to do this they could share with each other some portion
of the kept bits and see how many of them actually match. If the amount of
mismatched bits exceeds a certain number provided by models where Eve performs
optimally (see \cite{fuchs1997optimal}) then they should consider the channel to
be compromised and throw away all the bits: Every bit which Eve interacted with
has a chance to be corrupted and could contribute to the tally. If the number of
mismatches is low they can use the remaining bits as a random key.

Even though the protocol is unconditionally secure
\cites{shor2000simple,lo2001simple}, its physical implementation opens up
possibilities to attack the system \cite{scarani2014black}. In response to this
observation researchers started to work on device independent realisations of
these protocols. The last major section of this paper gives an overview of some very
recent results.

\section{More Quantum: BBM92, E91}

The BB84 protocol is based on Heisenberg's uncertainty principle (observation
effects the observed system) and the No-cloning theorem. There is another
feature of quantum mechanics that can help to make a quantum channel secure
beyond classical security.

Entanglement in quantum mechanics is an idea that
parts of a quantum system can be related beyond any classical connection. One
way to imagine this is to consider two coins that are connected in a
non-classical way as follows: If we flip one of the coins and get a heads then
flipping the other coins will result in getting a heads, too. The same is true
for the tails: If we flip one of the coins and get a tails then flipping the
other coin will result in a tails. The two coins either will show two heads or they will show two tails, never one heads and one tails. If we take two such coins, that is two coins
that are connected in such a non-classical way, far away from each other, they
will retain this property. This idea has bothered Albert Einstein and along with
two of his colleagues, Boris Podolsky and Nathan Rosen, suggested a thought  
experiment to point out that perhaps quantum mechanics is not complete. What was
then a thought experiment now a routine process in physics laboratories. A quantum bit pair that
possesses this kind of non-classical property is often called an EPR pair. 

In \cite{ekert1991quantum} Artur K. Ekert introduced a protocol, now referred to as E91,
that was based on entanglement. The method used Bell's theorem which  quantifies the
non-classical correlation resulting from entanglement and hence provides a way
to measure it.

The protocol works like this. A trusted provider of EPR pairs
sends one of the bits of this pair to Alice and the other bit of the same pair
to Bob. All the pairs are prepared in one of the four preferred states, for
example the bits are perfectly correlated. Using the dial analogy of the
previous section, if we looked at one of them and found that it points north/south
then measuring the other will reveal that it also points north/south. Or, if measuring
one of them showed that it is pointing east/west then measuring the other one
would show that it is pointing east/west, as well.

Both Alice and Bob have
three-three masks. For sake of simplicity we will refer to masks using only one of the directions of any pair of directions. Alice has one pointing east, one pointing north-east and one
pointing north, while Bob has one pointing north-east, one pointing north and
one pointing north-west. They choose a mask for each observation independently
from each other and from their previous choices. They perform the measurement on
the bit that they have received and record the result along with the mask they used. Once they are done with all
the planed measurements they announce their mask choices on a public channel
that is, again, susceptible to eavesdropping but resists all kinds of
tampering. They divide their results according to their mask choices: One group
will consists of those measurements where they used the same mask and the other
where they used different masks.

Next, they reveal their measurements results
for the second group, where they used different masks. Based on these results
they can compute the correlation between their measurements. If Eve tries to
interfere with the process by manipulating the entangled pairs, the computed
correlation will deviate from the number prescribed by quantum mechanics, hence
revealing her presence. 

In \cite{bennett1992quantum} Ch. H. Bennett, G. Brassard (of BB84 fame) and
N. David Mermin showed that the protocol (termed BBM92) works even without invoking Bell's
theorem: Eve cannot gain any information during the distribution of the
EPR pairs to Alice and Bob because information does not exist prior
measurement. Her interference would amount to modification of one of the quantum
bits of the pair. But that results in observable correlation loss and reveals
her presence.

One way for Eve to eavesdrop on the communication is to pose as an
EPR provider and secretly entangle the sent pairs with another quantum bit
available only to her. In the paper Bennett, Brassard and Mermin show that even
in this case Eve cannot gain any information without detection: In order to
avoid direct detection her quantum bit necessarily has to be disentangled (in
other words disconnected) from
the pair sent out to Alice and Bob. Any extra entanglement would show up in the
direct measurements.

Moreover, it turns out that the entanglement based protocol
BBM92 is equivalent to BB84 if the measurements are immediately performed, as
the quantum bits arrive, by at least one of
the communicating parties. In case of an entanglement based protocol, if the
measurement is postponed until the key is actually needed then any tampering
with the received quantum bits can be revealed later. For example, if a burglar
breaks into the office where the received bits are stored and changes them,
their act will be revealed during the measurement. This is not true for BB84
because the information stored by Alice is classical. 

\section{Device Independent Realisations}

As have been mentioned before, in \cite{scarani2014black} the authors point out
that since the quantum protocols are implemented using physical devices, these
devices are vulnerable to security threats. In case of an entanglement based
protocol it is possible to achieve security even if the device was manufactured
by an adversary as long as the device behaves as a true quantum mechanical
device: As it was mentioned in the previous section, the laws of quantum
mechanics prevent Eve, the adversary, from gaining information about the
communication in a stealthy way, that is without revealing her presence.

If Alice and Bob would like to use such a device to communicate, they have to
make sure that it behaves as a true quantum device. This can be achieved if the
device allows two kind of behaviours: key-generating mode and testing mode
\cite{arnon2018practical}. The communication between Alice and Bob is divided
into ``rounds'', each round could be a key-generating round or a testing
round. In one of the implementations (\cite{nadlinger2021device}), Bob decides
with certain probability if the round is a key-generating round or a testing
round and conveys his choice to Alice. In the key-generating round Alice and Bob
uses a protocol like E91 or a simplified version of it, to create bits for the
future key.

In the testing round they play a game. It is called the CHSH or
Clauser-Horne-Shimony-Holt game \cite{clauser1969proposed}. The game is played
by two cooperating players and there is a referee, usually called Charlie. The
players, Alice and Bob are not able to communicate during the game, but they can
agree on a strategy in advance and they can have a shared EPR pair since it is
not possible to communicate directly via the pair. Charlie chooses two numbers: He
chooses the first one to be $0$ or $1$ randomly with the same probability, and he
does the same for the second number. Then he sends the first number to Alice and the second
number to Bob. After receiving the number from Charlie, Alice responses with a
$0$ or a $1$ according to the chosen strategy and Bob does the same. Charlie
performs the logical AND operation on the two numbers he sent to the parties and
the addition modulo 2 operation on the received results. If the results are the
same Alice and Bob won.

In the classical local version of the game, when the
players don't use the EPR pair, they have $75\%$ at best to win the game. If
they use the EPR pair, by measuring it, they can increase this percentage to about
$85\%$. If the EPR pair is fake or compromised then their winning chance will be
less than $85\%$. Eve might be eavesdropping on their communication, but that
would be alright: In case of the CHSH game, there is a way to deduce at most how much
information is leaking to Eve, based on the reduction in winning probability. If
the amount is acceptable or they have ways to work around the problem
(e.g. privacy amplification, see \cites{nadlinger2021device,
  zhang2021experimental, liu2021high}) then they still can distill a key that is
unknown to Eve.

It seems, the researchers have finally overcome the challenge posed by the level
of currently available technology: The papers \cites{nadlinger2021device,
  zhang2021experimental, liu2021high} appeared at about the same time announcing
the implementation of truly device independent quantum key distribution. 

\section{Conclusion}

There are a couple of interesting questions that we could ask. What if quantum
mechanics is not always valid or there is some kind of post-quantum physics that
allows members of a more advanced civilisation to eavesdrop on us. The good news
is that, according to Jonathan Barrett, Lucien Hardy and Adrian Kent
\cite{barrett2005no}, all we need is a physics that does not allow faster than
light communication.

In this paper and throughout the literature (see e.g. \cite{ekert2014ultimate})
we use phrases like ``Alice chooses a representation'' or ``Bob chooses a
mask''. What if we cannot make free or independent choices? What if
superdeterminism is true and all our current choices are dependent or
correlated? Superdeterminism \cite{larsson2014loopholes} is a loophole in Bell's
theorem and of course it is not a new issue. John Bell already acknowledged it
and addressed the question in \cite{bell1995free}. The debate on this issue is
ongoing but seems to have little effect on the everyday physical models.

\section{Summary}

Quantum computers, when they arrive on the scene, will pose a real threat to
cryptographic protocols of today. It is good to know that the same technology
can arm us with solutions that can retain the security of our communication
networks which are so important in our lives.

\end{document}